# Magnetization reversal in sub-100nm magnetic tunnel junctions with ultrathin MgO barrier biased along hard axis

J.P. Cascales, D.Herranz[1], J.L. Sambricio[1], U. Ebels[2], J.A. Katine[3], and F.G. Aliev[1*]

(1) Dpto. Física Materia Condensada, CIII, Universidad Autónoma de Madrid, 28049, Cantoblanco, Madrid Spain

(2) SPINTEC, UMR 8191, CEA/CNRS/UJF, INAC, 38054 Grenoble Cedex, France

(3) Hitachi Global Storage Technologies, San Jose, CA 95135 USA

We report on room temperature magnetoresistance and low frequency noise in sub-100nm elliptic CoFeB/MgO/CoFeB magnetic tunnel junctions with ultrathin (0.9nm) barriers. For magnetic fields applied along the hard axis, we observe current induced magnetization switching between the antiparallel and parallel alignments at DC current densities as low as $4 \cdot 10^6 A/cm^2$. We attribute the low value of the critical current to the influence of localized reductions in the tunnel barrier, which affects the current distribution. The analysis of random telegraph noise, which appears in the field interval near a magnetization switch, provides an estimate to the dimension of the pseudo pinholes that trigger the magnetization switching via local spin torque. Micromagnetic simulations qualitatively and quantitatively reproduce the main experimental observations.

(*) corresponding author: farkhad.aliev@uam.es

Slonczewski's [1] and Berger's [2] prediction that a spin-polarized current between two ferromagnets could produce spin torque (ST), and in turn create steady magnetization dynamics or induce a magnetization reversal has been now widely confirmed experimentally. Spin torque magnetic random access memories (ST-RAM) [3] or microwave oscillators [4] are just two examples from a number of possible future spintronic devices based on spin torque. A key step towards the implementation of magnetic tunnel junctions (MTJs) as MRAM elements is the reduction of the critical currents/voltages needed for ST switching, in order to avoid heating and back-switching [5].

Another important field of potential applications of MTJs are MRAM-like biosensing chips based on arrays of magnetic sensors [6], scanning MTJ microscopy [7], etc. These new applications rely on the high external field sensitivity of MTJ sensors [8] when the external bias field direction is collinear with the hard axis (HA) [9]. In order to improve their spatial resolution, one should shrink these MTJ sensors in lateral size and consequently substantially decrease the MgO barrier thickness to keep the junction resistance reasonably small. This, however, may contribute to an enhanced influence of ST effects and of barrier inhomogenieties. With few exceptions [10], spin torque switching in HA biased MTJs remains poorly understood.

The first proposals to observe ST phenomena explored homogeneously distributed currents through wires or multilayer pillars. However, it was later found that a strongly non-uniform current flowing through a point contact with a diameter of a few tens of nm, created either mechanically [11] or lithographically [12], between two ferromagnetic films of large area separated by a *metallic* spacer may provide certain advantages for ST. For example, two or more closely situated contacts have been suggested to produce an effective phase lock-in of magnetization dynamics through spin wave exchange [13] or coherent vortex motion around contacts [14]. Such a lock-in is expected to increase the quality factor and power of the emitted microwaves.

Pinholes and barrier inhomogeneties are naturally present in magnetic tunnel junctions with ultrathin barriers [15] but they can also be created artificially by a soft breakdown [16]. The presence of pinholes was suggested to qualitatively modify the magnetoresistance [17] and the magnetization reversal mechanisms in MTJs, even in the absence of ST effects [18]. There has also been some controversy with respect to ST in MTJs with pinholes. While numerical calculations by Zhu [19], for single, and Meng [20], for multiple hot spots point to a decrease of the threshold ST current in comparison with non-broken MTJs, Finocchio et. al [21] predict an increase of the minimum current to excite microwave dynamics in junctions with pinholes.

Here we report on the experimental investigation, at room temperature, of current induced magnetization reversals (investigated via tunneling magnetoresistance and simulations) and of low frequency noise in CoFeB/MgO/CoFe/CoFeB MTJs of elliptical cross-section with major axes of sizes under 100nm, with low TMR (around 40%) and low resistance by area (RA) products of around $(2-4)\ \Omega \cdot \mu m^2$. This type of junctions (referred to as LTMR [4]) have been suggested to have localized reductions in the tunneling barrier, which could be described as an effective pinhole. We show that the fraction of area where the barrier is reduced may be estimated by analyzing the random telegraph noise (RTN) present in the samples. We find that due to inhomogeneous spin currents, the LTMR MTJs can be switched between the antiparallel (AP) and parallel (P) states using rather low current densities when magnetic fields are directed along the hard axis. Micromagnetic simulations, with an effective pinhole area close to what is estimated from the RTN, support the main experimental observations.

The multilayer nanopillars have the following structure: IrMn(6.1)/CoFe(1.8)/Ru/CoFeB(2)/MgO(0.9)/CoFe(0.5)/CoFeB(3.4) (the numbers indicate de thickness of the layer in nm) and have been fabricated by Hitachi Global Storage Technologies. The pinned layer consists of two ferromagnetic layers which are antiferromagnetically (AF) coupled through a thin ruthenium layer. Another AF layer, exchange-coupled to the lower ferromagnetic layer, is added to rigidify the synthetic AF. The MgO barrier is deposited by sputtering and the free layer consists of a bi-layer of CoFe/CoFeB. The measured nanopillar devices have elliptical cross-sections of different sizes, with the minor and major axes ranging from 40×80 to 65×130 (in nm). The easy axis (EA) direction is parallel to the pinned layer's magnetization and it coincides with the major axis of the ellipse, while the in-plane hard axis (HA) is perpendicular to the easy axis. The devices are embedded in impedance matched RF coplanar waveguides for electrical contacting using special RF probes. The devices were biased by a DC current. Here, positive currents mean electrons tunneling from the pinned to the free layer and negative currents vice-versa. For the low frequency noise measurements, the voltage across the device was pre-amplified by home-made amplifiers and measured with a nanovoltmeter. The dc component of the voltage was filtered out and the ac fluctuations amplified by an SR560 commercial amplifier. The pre-amplified fluctuations were processed by a SR785 spectrum analyzer (up to 102.4 kHz) to obtain the noise spectra. The results shown in this letter belong to a 40x80nm ellipse, although the same behavior was observed in several other devices (of similar and different sizes).

When no external field is applied, the free electrode's magnetization ($M_{free}$) and the pinned electrode's magnetization ($M_{pinned}$) are aligned parallel. This is the state of lowest resistance, called the parallel or P state. The highest resistance of the sample is reached when H is applied along the EA and the angle between the moments of the electrodes is 180º. This is known as the antiparallel or AP state. Now, if a high enough field is applied along the HA (around 1.5kOe, Fig. 1(a)), $M_{free}$ fully aligns with H. Since $M_{pinned}$ remains fixed, for this saturation field the angle between the magnetization of the ferromagnetic electrodes is around 90º. We will call this state $AP_{HA}$, the hard axis anti-parallel state. Then we have that $R(P)<R(AP_{HA})<R(AP)$.

The TMR curves shown in Fig. 1(a) were obtained when the field applied along the HA was swept from a high positive to a high negative value. For positive currents, the sample behaves normally, and reaches the P state when the field approaches 0. However, for negative currents, when the field is lower than the switching field, i.e. $|H|<H_{switch}$, ST effects overcome the external H and switch the sample to a high resistance state. The resistance of the sample in this state is higher than in the $AP_{HA}$ state, hence we argue that the sample switches to the AP state. Figure 1(b) shows a phase diagram of the magnetic state of the sample, dependent on H and the applied current, constructed from TMR curves at different currents. It can be seen that a region appears at some negative critical current density where the sample is switched to the AP state.

We imagine the switching process as a double well potential, where for $|H|>H_{switch}$, the minimum of energy corresponds to the typical HA TMR configuration and for $|H|<H_{switch}$, the minimum of energy corresponds to the AP state due to ST. In the vicinity of $H_{switch}$ we get a bi-stable situation and thermally activated RTN is detected. The conditions for such a bi-stability should primarily exist in the areas of the soft magnetic electrode which are located close to the pinholes, i.e. where the current density is the highest.

We have used the low frequency noise (LFN) measurements as a tool to quantify the barrier and current inhomogeneities in these MTJs with ultra thin barriers. Special attention has been paid to random telegraph noise as a potential source of useful information for estimating the size of the "defective" region of the barrier. The spectra in these samples usually present 1/f noise, save for the fields where RTN is present. The 1/f noise is quantified by the Hooge factor $\alpha$, which is obtained from the relation $S_v = \alpha \cdot V^2/(A \cdot f^\beta)$, with V the applied voltage, A the area of the sample and the exponent $\beta$ which is a fitting parameter [22]. The field dependence of the Hooge factor revealed a clear maximum in noise centered around the field value where the resistance switch takes place. Also, the exponent $\beta$ goes to 0 for these same field values, i.e. the curve becomes flat and Lorentzian-like. Neither $\alpha(H)$ nor $\beta(H)$ are shown for briefness' sake. These features clearly show the range of H which presents RTN.

The characterization of the RTN was then carried out by analyzing the spectra and time-series at these fields, following the method explained further below. Figure 2 shows typical RTN features in the LTMR samples, where the field is directed along the HA. Fig. 2(a) shows typical time series of the voltage fluctuations for magnetic fields in and outside the field range where a strong RTN is detected (H=+1700 and +600 Oe). On the other hand, the graph corresponding to the reorientation transition ( H=+190 Oe) shows well defined step-like jumps between two voltage levels. The amplitude of these RTN fluctuations is a factor of $10^2$ larger than RTN for higher fields. Figure 2(b) shows that for the field values in the range where the magnetization reversal takes place (H=+190 Oe) one observes a Lorentzian-like spectrum, typical of RTN. On the other hand, outside the range (H=+1700 Oe and H=+600 Oe), the spectrum is textbook 1/f.

The graphs shown in figures 2(a) and (b) correspond to $J=-1.2\cdot 10^7 A/cm^2$. For each current, the fluctuating magnetic moment ($\Delta m$) involved in the RTN is estimated in the following way. The fluctuating voltage ($\Delta V$) is obtained from the time-series as the difference in voltage between the two levels ("up" or "down"), by fitting two Gaussians to the histogram of the time-series. The spectra are fitted by the theoretical curve $S_V = \Delta V^2/(t_{up}+t_{down}) T^2/(1+(2\pi T f)^2)$ [23], where $T^{-1}=1/t_{up}+1/t_{down}$. Then we obtain $\ln(t_{up}/t_{down})$ with respect to H for a specific current, and we fit this by the Arrhenius law $\ln(t_{up}/t_{down})=C+2\Delta m H/k_B T$ (where C is a constant) as shown in Fig. 2(c). Figure 2(d) shows the estimation of $\Delta m$ for different currents, which is found to be around $(2-4)\cdot 10^5 \mu_B$, for both resistance switches (AP-P and P-AP) present in each curve. Similar effects were observed in several of the junctions. Considering the moment per atom in CoFeB to be of $1\mu B$ [24], its lattice parameter $a=0.284nm$, an fcc structure (hence there are $4\mu B$ in a volume of $a^3$) and if we suppose that the fluctuating moment is only present in the free layer, then we estimate that the volume which corresponds to $\Delta m=4\cdot 10^5 \mu_B$ is 23% of the volume of the free electrode, which fluctuates and generates the RTN. This fraction is obtained from $V_{RTN}/V_{electrode}$ where $V_{RTN} = \Delta m \cdot a^3/4\mu B$ and $V_{electrode}=\pi xyz$, where $x=20nm$, $y=40nm$ and $z=3.9nm$.

    We have carried out numerical simulations, with OOMMF [25], of junctions with and without pinholes described by a simple qualitative model in order to account for the observed phenomena. The reduction of the barrier or the presence of a pinhole are modeled as a region of area *a* in the insulating barrier which concentrates the current going through the structure, as schematically shown in Fig. 3(b). If J is the current density flowing through the electrodes of area A, then the current density in the pin-hole is $J\cdot A/a$. For negative currents, electrons flow from the free to the pinned layer and this favors an AP alignment of the electrodes' magnetizations. The constants for CoFeB

used in the simulations are: spin polarization P=0.5 and saturation magnetization $M_s$=1150kA/m. The TMR curves (Fig.3(a),(b)) have been calculated by obtaining the average angle between the free and pinned layer's magnetizations, and using the expression R(θ)=R(0)+ΔR·(1-cosθ)/(2+χ(1+cosθ)) [26]. Since H was only applied along the HA in this sample, we have to estimate ΔR. From our data, TMR(π/2)=23% and R(0)=945Ω, so with χ=0.5 we obtain that ΔR =0.6·R(0), i.e. a reasonable TMR=60%.

The results of the simulation (Figure 3) closely resemble the experimental results with the exception of the AP→P switch for negative fields. Our simulations show that a smaller current is necessary to switch the free layer's magnetization if the current through some region of the barrier, where there is a higher effective current density, compared to a perfect junction. Fig. 3(a) shows that a perfect barrier needs a much higher current density to obtain a resistance switch (J=-2·10$^7$A/cm$^2$) than junctions with local barrier reductions, as seen in Fig. 3(b). Figures 3(c) and (d) show a phase diagram constructed from all the TMR curves for an MTJ with (d) and without (c) a pinhole. Figure 3(d) closely resembles the experimental data shown in Fig. 1(b). Indeed, for the ratio A/a=5.3 (for which close to 20% of the insulating surface concentrates current), we obtain a switch to the AP state for low H at precisely J=-4·10$^6$ A/cm$^2$, as can be seen in Fig. 3(b) and (d). If the fluctuating moments are located in the free layer, and strictly above the area *a* of the pinhole, this corresponds to 20% of the volume of the free layer, which is remarkably close to the 23% mentioned above.

Some disagreement between simulations and experiment could be related to the fact that the simulations are done at zero temperature while the experiments were carried out at 300K. Besides, one should not completely exclude some contribution of electric origin to the observed RTN (see Fig. 2a). The scenarios which could describe the barrier reduction in our MTJs are: (a) a reduction of the effective barrier (pseudo pinhole) which involves direct tunneling or (b) centered or off-center defect states in the barrier which induce sequential tunneling, appearing as electric RTN. The following arguments disprove electric RTN as the main source of the random telegraph noise: RTN of a purely electric origin observed in sub 100nm MTJs with ultrathin (<1 nm) MgO tunnel barriers, showed to be field independent [27], and it appears for field values outside the magnetization reversal range as in our experiment (Fig.2a,b). Field dependent RTN appears for a range around the magnetization reversal and is two orders of magnitude higher in amplitude than the electric RTN which is detected at higher fields. Moreover, RTN due to domain walls or magnetic inhomogenieties [28] should also be excluded since the lateral dimensions of the MTJs under study are smaller than 100nm, which is below the typical DW width. In order to evaluate more precisely what the contribution of electric RTN is, shot noise measurements should be carried out. Direct (indirect) tunneling should give Poissonian (sub-Poissonian) shot noise [29,30].

Finally, we remark that qualitatively different low frequency noise was observed in high TMR (above 70%) junctions (called HTMR [4]). These MTJs, expected to have a more uniform, pseudo-pinhole free barrier, revealed a decrease in the Hooge factor with an increasing applied bias, similarly to what was previously observed for Fe/MgO/Fe MTJs with 2-3 nm thick MgO barriers [31-33].

In conclusion, a detailed investigation of magnetoresistance, spin torque switching and random telegraph noise in sub-100nm MTJs with an external magnetic field applied along the hard axis was carried out. The experimental conclusions are supported by micromagnetic simulations, which show that local reductions of the MgO barrier could be responsible for the substantial decrease in critical current needed for spin torque induced magnetization switching. The obtained results should help to define the "current window range" for the potential application of nm sized magnetic tunnel junctions as ultra small field sensors.

Authors acknowledge G.Finocchio for discussions. The work has been supported by the Spanish MINECO (MAT2012-32743, CONSOLIDER CSD2007-00010) and Comunidad de Madrid (P2009/MAT-1726) grants.

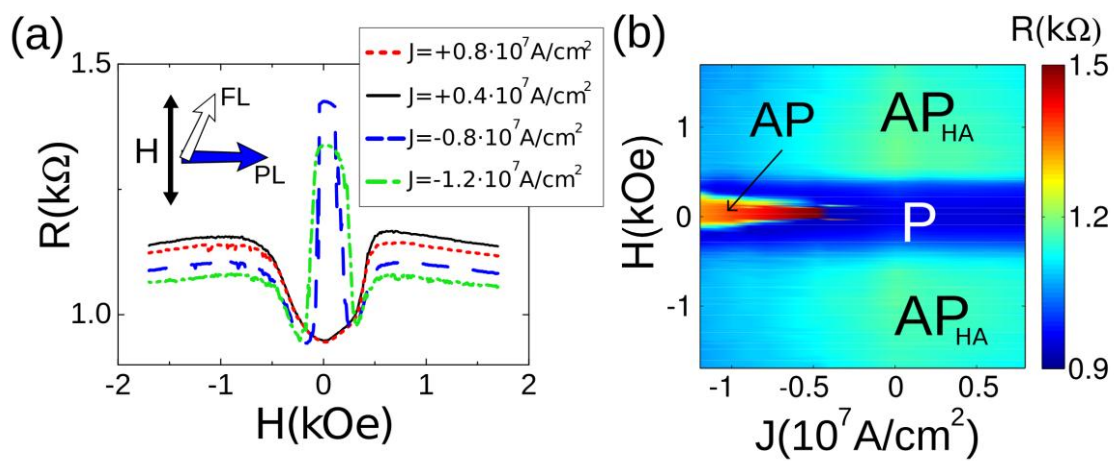

Figure 1. (a) TMR curves for several currents with H in the hard axis direction. Positive currents show a typical TMR curves, while negative currents show an abrupt increase in resistance at low fields. (b) Phase diagram of the magnetic state of the hard axis sample, constructed from the TMR curves. At low H, the switch to the AP state occurs at around $J=-4\cdot 10^6$ A/cm$^2$.

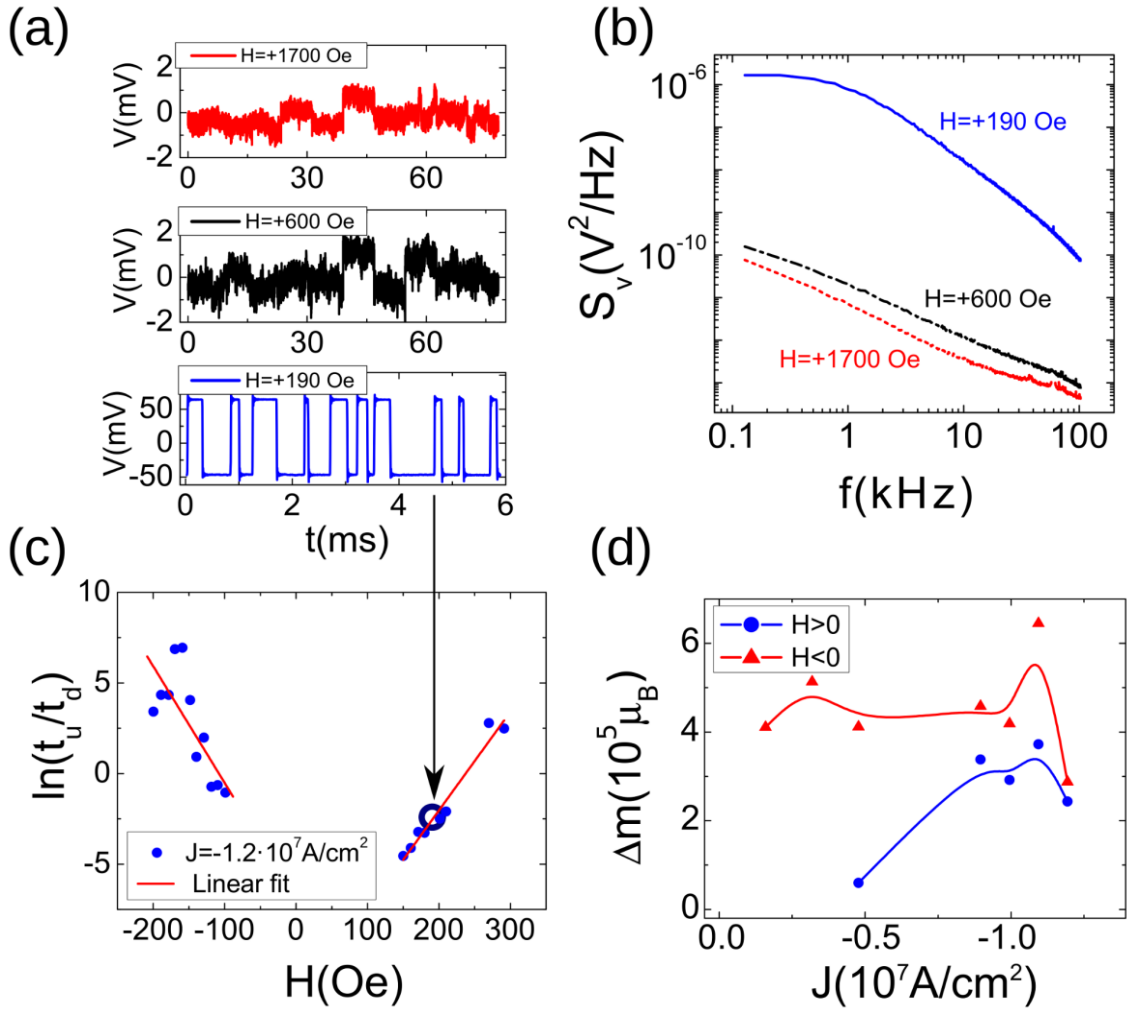

Figure 2. (a) Voltage fluctuations for three different applied magnetic fields for J=-1.2·$10^7$A/cm$^2$. The field range around the magnetization reversal (H=+190 Oe) presents RTN fluctuations which are two orders of magnitude higher in amplitude than other values of the field (H=+1700 Oe, H=+600 Oe). (b) Lorentzian-like spectrum typical of RTN at H=+190 Oe and a typical 1/f spectrum for H=+1700 Oe and H=+600 Oe. (c) Linear fit to an Arrhenius-type law of the $t_{up}/t_{down}$ ratio, from which the fluctuating moment for each current is estimated. (d) Estimation of the fluctuating moment Δm with respect to the applied current. The Δm obtained corresponds to the fluctuation of around a fifth of the volume of the free layer.

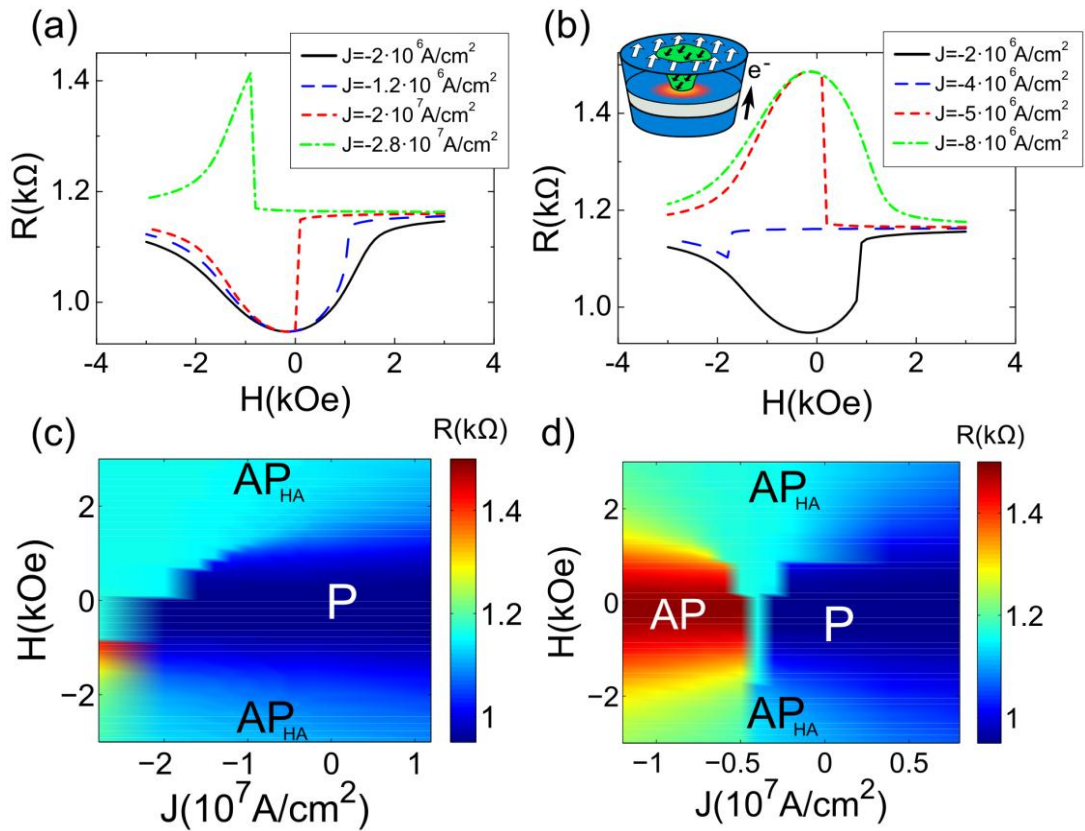

Figure 3. (a) Simulated TMR curves for different current values in a junction with a perfect barrier. (b) Simulated TMR curves for different currents with a pin-hole in the barrier. Lower current values are needed to switch the free layer toward an AP alignment than for the pin-hole free structure. (c) Phase diagram constructed from the TMR curves of the perfect junction. Current densities of around $J=-2\cdot 10^7 A/cm^2$ are needed to switch the free layer to an AP state. (d) Phase diagram constructed from the TMR curves of the junction with a pin-hole. The current density needed to switch to the AP state, much lower than for a perfect barrier, corresponds to the experimental result $J=-4\cdot 10^6 A/cm^2$ for A/a=5.3.